# Measurements of quasi-particle tunneling in the $\upsilon = 5/2$ fractional quantum Hall state

X. Lin,[1, *] C. Dillard,[2] M. A. Kastner,[2] L. N. Pfeiffer,[3] and K. W. West[3]

1 International Center for Quantum Materials, Peking University, Beijing, P.R. China 100871

2 Department of Physics, Massachusetts Institute of Technology, Cambridge, Massachusetts 02139, USA

3 Department of Electrical Engineering, Princeton University, Princeton, New Jersey 08544, USA

*Abstract:*

*Some models of the 5/2 fractional quantum Hall state predict that the quasi-particles, which carry the charge, have non-Abelian statistics: exchange of two quasi-particles changes the wave function more dramatically than just the usual change of phase factor. Such non-Abelian statistics would make the system less sensitive to decoherence, making it a candidate for implementation of topological quantum computation. We measure quasi-particle tunneling as a function of temperature and DC bias between counter-propagating edge states. Fits to theory give $e^*$, the quasi-particle effective charge, close to the expected value of e/4 and $g$, the strength of the interaction between quasi-particles, close to 3/8. Fits corresponding to the*

*various proposed wave functions, along with qualitative features of the data, strongly favor the Abelian 331 state.*

## I. INTRODUCTION

The collective interactions of a two-dimensional electron gas (2DEG) in a strong perpendicular magnetic field $B$, give rise to the fractional quantum Hall effect (FQHE).[1] Because of the energy gap in the bulk,[2] motion of the quasi-particles that arise in the FQHE is generally constrained to one-dimensional chiral edge channels. However, if two opposite channels are brought close together, quasi-particles may tunnel between them. Studies of such tunneling have led to measurements of the quasi-particle charge[3, 4] and creation of quasi-particle interferometers.[5, 6]

The states comprising the FQHE are determined by the filling factor $\upsilon = n/(B/\Phi_0)$, where $n$ is the electron sheet density and $\Phi_0 = h/e$ is the quantum of magnetic flux. The $\upsilon$ = 5/2 state[7] is of particular interest because it is one of only a few physically realizable systems thought to possibly exhibit non-Abelian particle statistics.[8-13] A number of different ground state wave functions have been proposed for the 5/2 state, some with non-Abelian statistics and some with prosaic Abelian statistics. Were the existence of non-Abelian statistics confirmed, it would be an exciting discovery of a new state of matter and would possibly enable topological quantum computation.[14] A great deal of theoretical and experimental work has been done on the 5/2 state recently.[15-32] Experimentally, the quasi-particle charge $e^*$ has been found to be consistent with the predicted value e/4.[24, 25, 31] Numerical simulations indicate a preference for the non-Abelian Pfaffian and anti-Pfaffian wave functions over various Abelian wave functions.[18, 21, 23, 32, 33] The degree of electron spin polarization also provides valuable information about the wave function, but experimental results are contradictory.[29, 30] Recent experimental results from an interferometer have been interpreted as evidence for non-Abelian statistics at $\upsilon = 5/2$.[26, 27] The observation of a counter-propagating neutral mode is also most easily explained by the existence of a non-Abelian state.[28]

We have studied the $\upsilon = 5/2$ state in two different quantum point contact (QPC) geometries, and present temperature and DC bias dependence of quasi-particle tunneling conductance across each QPC in the weak tunneling regime. We have improved the signal-to-noise ratio by a factor ~2 compared to previous similar measurements.[24] By fitting these results to the theoretical form,[34] we extract the quasi-particle charge $e^*$ and interaction parameter $g$. The resulting $e^*$ is in agreement with the predicted value of e/4, and the value of $g$ best agrees with that predicted for the Abelian 331 state. Fixing $g$ at the values predicted by other proposed states produces fits that are qualitatively and quantitatively worse. In addition, qualitative features of the DC bias dependence also favor the 331 state.

## II. EXPERIMENTAL DETAILS

The device used is the same as one of those studied by Radu et al.,[24] and we briefly summarize its characteristics here. The GaAs/AlGaAs heterostructure has a measured mobility of $1 \times 10^7$ cm$^2$/V s and electron density $n = 2.6 \times 10^{11}$ cm$^{-2}$. The mobility is only half that previously reported, but the cause of this degradation is unknown. The sample still exhibits a strong 5/2 fractional quantum Hall effect, with a quantized Hall plateau and vanishing longitudinal resistance.[35] Metallic top gates are biased negatively to deplete the underlying two-dimensional electron gas and induce tunneling between edge channels. The gate pattern is shown in Fig. 1(a). Two different QPC geometries are created by applying negative voltages to some of the gates while keeping the remaining gates grounded. Geometry A is a short QPC of nominal width ~0.6 µm, and Geometry B is a long channel of nominal width ~1.2 µm and length ~2.2 µm. For convenience we refer to both geometries as QPCs. The gates biased to form the two geometries are listed in the caption of Fig. 1. The measurement setup is illustrated in Fig. 1(b). A DC current $I_{DC}$ of up to ±10 nA with a 0.4 nA AC modulation is applied to the source at one end of the Hall bar, with the drain at the other end. Using standard lock-in techniques at 17 Hz, measurements are made of the differential Hall resistance ($R_{XY}$) and longitudinal resistance ($R_{XX}$) at points remote from the QPC and of the differential diagonal resistance ($R_D$) across the QPC. Experiments are made in a dilution refrigerator with a mixing chamber base temperature of ~5 mK. The temperatures quoted below are electron temperatures, which track the mixing chamber temperature well down to ~20 mK. Lower electron temperatures are calibrated against thermally broadened Coulomb blockade peaks in a quantum dot and against quantum Hall features showing linear temperature dependence.[35]

In order to preserve the same electron density, and hence filling factor, both inside and outside the QPC, we anneal the device with bias voltage applied to the gates at 4 K for approximately 60 hours before cooling to base temperature.[24] Each geometry is annealed by applying -2400 mV to the gates listed in the caption of Fig. 1, while leaving the other gates grounded. After annealing and lowering the temperature, the gate voltage $V_G$ is constrained to the range of -2400 mV to -1800 mV. We find that after annealing the filling factor in the QPC matches that of the bulk of the Hall bar. This is important in order to be confident that the measurements reflect tunneling of $\upsilon = 5/2$ quasi-particles from one chiral edge state to the other. The success of annealing is confirmed by the observation that $R_{XY}$ (sensitive to the 2DEG far from the QPC) and $R_D$ (sensitive to the QPC) exhibit the same dependence on magnetic field.[35] In particular, the integer QHE plateaus begin and end at the same values of magnetic field, and the low field slopes of $R_{XY}$ vs B and $R_D$ vs B are the same. This identical device in Geometry B has been previously measured in the strong tunneling regime[24] and we have repeated these measurements with similar results. Here we anneal at a less negative gate voltage enabling us to access the weak tunneling regime to much lower temperature.

## III. RESULTS

For each geometry we measure the DC bias and temperature dependence of $R_D$ at $\upsilon = 5/2$, as shown in Fig. 2(a), (b). The peak positions deviate from $I_{DC} = 0$ by ~0.2 nA because of hysteresis in the sweep direction; all measurements shown in this paper are taken with increasing DC bias, and the small offset is subtracted when fitting. The magnetic field and gate voltage are chosen, following the measurement technique described in Radu et al.,[24] to maximize the temperature range exhibiting a zero-DC-bias peak and to minimize variations in the background resistance.[35] The magnetic field is set to 4.31 T, which is the center of the $R_{XY}$ plateau for $\upsilon$ = 5/2. Geometry A is measured at $V_G$ = -2100 mV and Geometry B at -2148 mV.

In the limit of weak quasi-particle tunneling, $R_D$ is linearly related to the differential tunneling conductance $g_T$ by

$$g_T = (R_D - R_{XY})/R_{XY}^2, \tag{1}$$

provided that the sample is in an integer or fractional quantum Hall state. This allows us to fit to the theoretical form for weak quasi-particle tunneling in an arbitrary FQHE state[34]:

$$g_T = AT^{(2g-2)} F(g, \frac{e^* I_{DC} R_{XY}}{k_B T}). \tag{2}$$

The quasi-particle charge $e^*$ and interaction parameter $g$ are physical constants characterizing the FQHE state, while the fit parameter $A$ accounts for the tunneling amplitude. $F$ has a complicated functional form, peaked at zero DC bias.[35] $F$ approaches zero at infinite DC bias and for $g < 1/2$ exhibits minima on the sides of the zero-bias peak; these minima are absent for $g \geq 1/2$. Equation (2) predicts that the zero-bias peak height follows a power-law temperature dependence and that the full-width at half maximum in DC bias is linear in temperature. Like Radu et al.[24] and as seen in Fig. 2(a) and (b), we find a background resistance $R_\infty$, which is larger than the expected quantized value 0.4 $h/e^2$. However, this background is independent of DC bias and temperature within the resolution of the measurement, under the measurement conditions chosen.

A least-squares fit to Equation (2) of the $R_D$ measurements for Geometry B for many temperatures is shown in Fig. (2c). We limit the temperature range for fitting to 20–80 mK for Geometry A and to 27–80 mK for Geometry B to ensure that only the weak tunneling regime is included. At lower temperatures the DC bias curves start to exhibit features associated with strong tunneling.[24, 36, 37] However, we note that the fitting results do not change significantly if a different temperature range is chosen. Within the chosen range, measurements for all temperatures are fit *simultaneously* with the same fitting parameters. The best fit for Geometry A returns $e* = 0.25e$ and $g = 0.42$ and for Geometry B returns $e* = 0.22e$ and $g = 0.34$. These values are similar to those obtained in a previous weak tunneling measurement[24] ($e* = 0.17e$, $g = 0.35$), but we find $e*$ much closer to the predicted value of $e/4$.[38]

In order to understand the level of confidence we can place in these fit results, we fix ($e*$, $g$) pairs over a range of values and fit to the weak tunneling form, allowing $A$ and $R_\infty$ to vary. The residual of each fit is divided by the measurement noise of $\sim 2 \times 10^{-4}$ $h/e^2$ and plotted against the ($e*$, $g$) pairs in Fig. 3. Pairs of ($e*$, $g$) values with a fit residue no larger than the noise fall within the "1" contour. Also indicated are the pairs of ($e*$, $g$) values corresponding to the proposed wave functions listed below.

## IV. INTERPRETATION AND DISCUSSION

The simplest interpretation of our data is that the value of $g$ derived from the fits directly reflects the nature of the quasi-particles, and therefore the wave function, of the 5/2 state in our system. This allows us to distinguish between competing proposals by their expected value of $g$. Proposed Abelian states include 331[39, 40] with $g = 3/8$, and K = 8[41] with $g = 1/8$. Non-Abelian states include the Pfaffian[9] with $g = 1/4$, the particle-hole conjugate anti-Pfaffian[11, 12] with $g = 1/2$, and $U(1) \times SU_2(2)$[10] with $g = 1/2$. All these states support quasi-particles both of charge e/4 and charge e/2. However, tunneling measurements are expected to be dominated by quasi-particles of charge e/4.[42] Hence we examine the fits to Equation (2) with $e^*$ fixed at e/4 and $g$ fixed at a value predicted by one of the proposed states: 1/8, 1/4, 3/8, or 1/2. Of these four options, fits with $g = 3/8$ and $g = 1/2$ produce the lowest residuals, as can be seen in Fig. 3; fits for $g = 1/8$ and 1/4 are very poor, both quantitatively and qualitatively.[35]

Fits for Geometry A with $g = 1/2$ and 3/8 are shown in Fig. 4(a) and (b) respectively, and similarly for Geometry B in Fig. 4(c) and (d). The fits with $g = 3/8$, corresponding to the Abelian 331 state, are clearly the best, following the data closely, including the temperature dependence of the peak height and the obvious minima on either side of the main peak. For $g = 1/2$, the minima on either side of the peak are absent, and peak heights are not well described. Our ability to better discriminate between $g = 1/2$ and $g = 3/8$ than with previous weak tunneling measurements[24] results from a factor of about two reduction in the noise. Reaching a lower level of noise is particularly important because it makes the DC bias minima, which are important qualitative features of equation (2), clearly distinguishable from the noise. The minima are even more prominent at base temperature, as can be seen in Fig. 2.

In fact, the presence of these minima, which have also been observed in previous strong tunneling measurements,[24] is an indication that g is strictly less than 1/2. Expanding the exact form of the current transmitted through a QPC in the FQHE around the high DC bias limit[36] one finds

$$\frac{1}{R_D} = \frac{dI}{dV_D} = g + C\left(\frac{T_B}{V_D}\right)^{2(1-g)} (2g-1) + \cdots . \tag{3}$$

Here, $C$ is a negative constant, $T_B$ is a temperature scale reflecting the strength of the edge channel interaction in the QPC, and $V_D$ is the diagonal voltage. As $T_B/V_D$ increases (corresponding to $|I_{DC}|$ decreasing from infinity), $R_D$ decreases for $g < 1/2$, is constant for $g = 1/2$, and increases for $g > 1/2$. Since $R_D$ eventually increases as the bias approaches zero, this produces a minimum in $R_D$ only for $g < 1/2$. This behavior is also reflected in the weak tunneling formula, Equation (2). This analysis neglects higher order terms, which could conceivably produce a minimum in $R_D$ at $g = 1/2$, but to lowest order the presence of minima requires $g < 1/2$.

It may be that the value of $g$ extracted from Equation (2) does not directly reflect the nature of the 5/2 state in our system. Reconstruction of edge channels[43] and interactions between edge channels may cause the value of $g$ observed by tunneling experiments to change.[44, 45] The presence of striped phases, which have been found to be energetically favorable for some confinement potentials,[17] or random domains of different ground state wave functions[15] may also complicate matters. Even the Luttinger liquid theory of quasi-particle tunneling may be an incomplete description of experimental systems. Experimental results of electron tunneling in cleaved edge overgrowth samples are not fully described by the predictions of chiral Luttinger liquid theory, indicating that perhaps a more nuanced picture is needed to accurately describe tunneling experiments.[46] Measurements of transmission through QPC devices at in the lowest Landau level provide some confidence that Luttinger liquid theory can be applied to quasi-particle tunneling across a QPC.[37, 47, 48] Ideally, one would like to repeat the measurement technique of Radu et al.[24] at a simple filling fraction, such as $\upsilon = 1/3$, and fit the results to equation (2). Unfortunately, the sample studied here is not suitable for such measurements, because of the relatively wide width of the QPC and the low electron density.

## V. CONCLUSION

In conclusion, fits based on quasi-particle weak tunneling theory[34] favor the presence of the 331 Abelian wave function in our sample, while excluding other states, including the non-Abelian Pfaffian and anti-Pfaffian. Particularly telling is the presence of minima in the DC bias dependence, which requires $g < 1/2$. Given previous studies favoring the Pfaffian and anti-Pfaffian wave functions[18, 21, 23, 29, 32, 33] or a non-Abelian wave function in general,[27, 28] it seems possible that different states may be physically realizable at $\upsilon = 5/2$. The device geometry and heterostructure characteristics may be important factors in determining which state is favored. For example, there is numerical evidence that the strength of the confinement potential influences the wave function exhibited in a FQHE system at $\upsilon = 5/2$.[17] Better understanding of the effects of these factors on the 5/2 state will likely be vital for any efforts to further explore non-Abelian particle statistics or realize a topological quantum computer. We recommend that similar measurements of $e^*$ and $g$ be performed on other heterostructures and device geometries, especially the ones in which evidence of a non-Abelian wave function has been observed.[27, 28]

We are grateful to Claudio Chamon, Dmitri Feldman, Paul Fendley, Charles Marcus, Chetan Nayak, and Xiao-Gang Wen for helpful discussions. We thank Jeff Miller from the Marcus Lab for sample fabrication at Harvard's Center for Nanoscale Systems, with support from Microsoft Project Q. The work at MIT was funded by National Science Foundation under Grant No. *DMR-1104394*. The work at Princeton was partially funded by the Gordon and Betty Moore Foundation as well as the National Science Foundation MRSEC Program through the Princeton Center for Complex Materials (*DMR-0819860)*.

References:

* linxi07@gmail.com

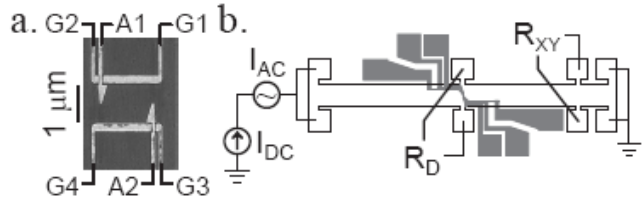

FIG. 1. Device image and measurement setup.

(a) Scanning electron micrograph of a device fabricated similarly to the one used in this experiment. Gates A1, G3, and G4 are biased negatively for Geometry A, with the other gates grounded. For Geometry B: G1, G2, G3, and G4 are energized, with A1 and A2 grounded. (b) Simplified diagram of the Hall bar mesa and measurement setup. The mesa is outlined and top gates are shaded grey.

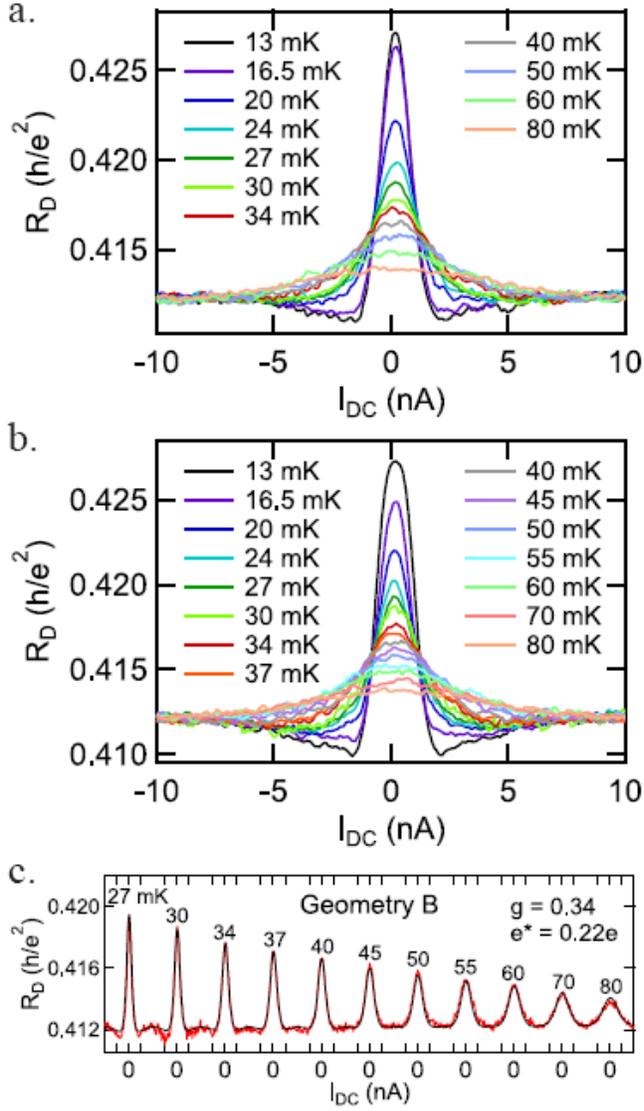

FIG. 2. (Color) DC bias and temperature dependence of $R_D$, with best fit.

(a) $R_D$ measured in Geometry A with an applied gate voltage of $V_G$ = -2100 mV. (b) $R_D$ measured in Geometry B with $V_G$ = -2148 mV. For both, the magnetic field is set to the center of the $\nu$ = 5/2 $R_{XY}$ plateau (B = 4.31 T) and the sample is annealed at $V_G$ = -2400 mV as discussed in the text. (c) Least-squares best fit of $R_D$ to Equation (2) for Geometry B, with all temperatures in the range 27–80 mK fit simultaneously. Tunneling conductance peaks at each temperature (labeled on graph) are concatenated to produce a single curve. Experimental results are red and the fit is black. Ticks indicate 0 and ± 5 nA for each temperature.

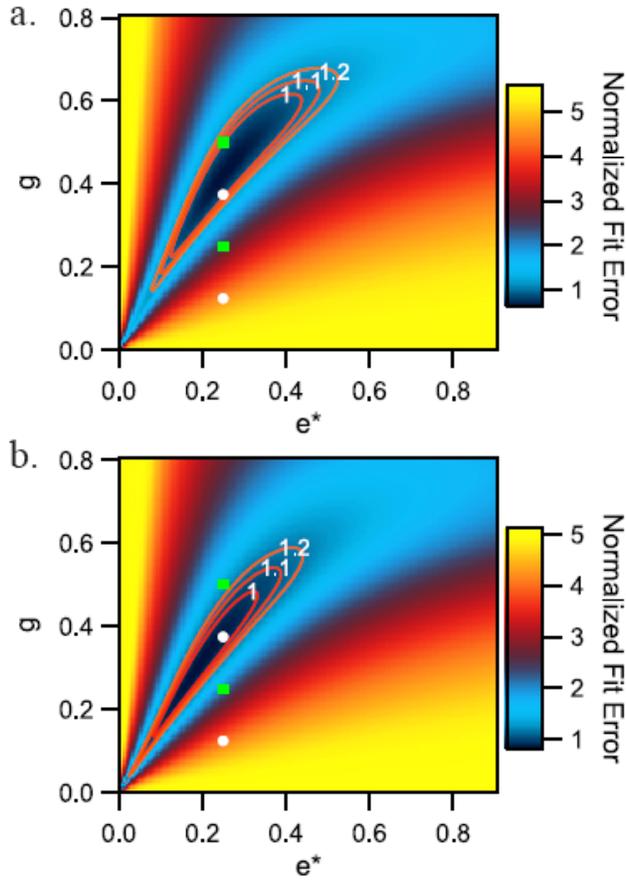

FIG. 3. (Color) Matrix of fit residual for fixed pairs of ($e^*$, $g$) divided by the experimental noise.

(a) Results for Geometry A. (b) Results for Geometry B. For both, pairs of ($e^*$, $g$) corresponding to proposed non-Abelian states (green squares) and Abelian states (white circles) are plotted; see the text for more details. Contours of the normalized fit error are included as guides to the eye.

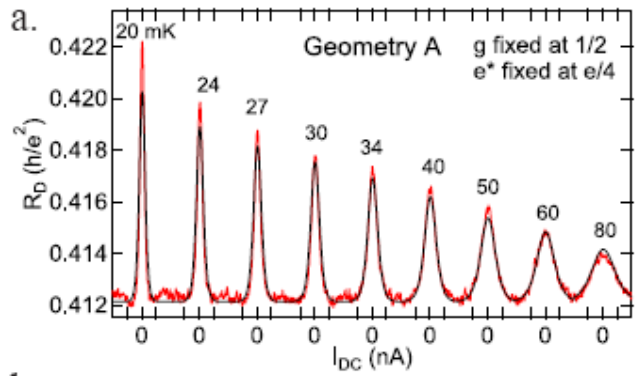
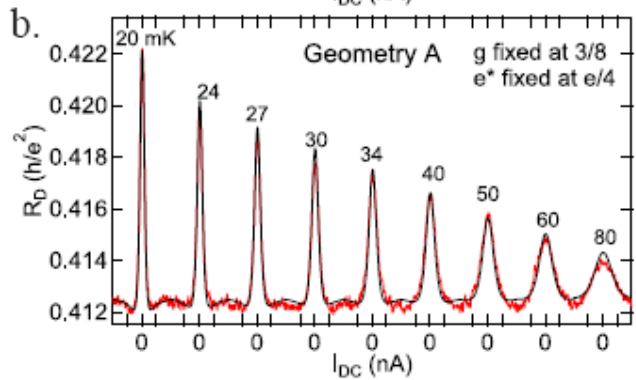
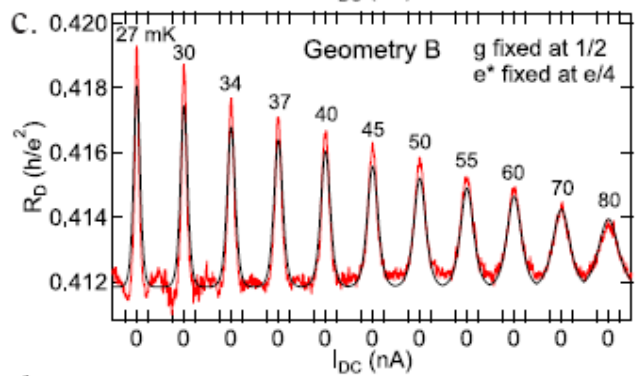
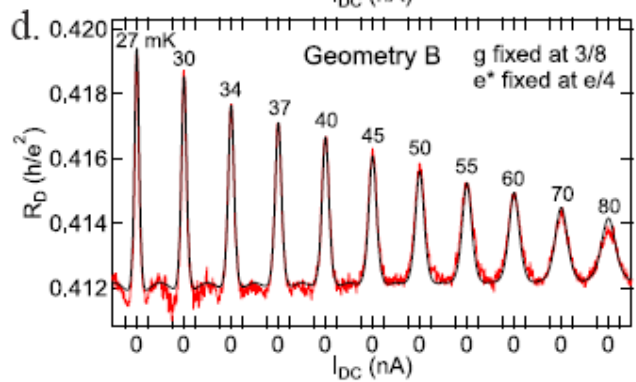

FIG. 4. (Color) Fits of $R_D$ to the theoretical form of quasi-particle weak tunneling with fixed $e*$ and $g$.

(a) and (b) Fits of $R_D$ in Geometry A to Equation (2) with fixed $g = 1/2$ (anti-Pfaffian and $U(1) \times SU2(2)$ wave functions) and $g = 3/8$ (331 wave function), respectively. (c) and (d) Fits of $R_D$ in Geometry B to Equation (2) with fixed $g = 1/2$ and $g = 3/8$, respectively. In all cases $e*$ is fixed to $e/4$ and all temperatures shown are fit simultaneously with the same fit parameters. Tunneling conductance peaks at each temperature (labeled on graph) are concatenated to produce a single curve. Experimental results are red and fits are black. Ticks indicate 0 and ± 5 nA for each temperature.